# ON THE GEOLOGICAL TIME EVOLUTION OF VOLCANISM IN THE INNER SOLAR SYSTEM

Varnana.M.Kumar [1], T.E.Girish[2*], Thara.N.Sathyan[1], Biju Longhinos[3], Anjana A.V.Panicker[4] and J.Binoy[1]

[1]Department of Physics, Government College for Women, Thiruvananthapuram, Kerala, India 695014
*E mail: tegirish5@yahoo.co.in
[2] Department of Physics, University College, Thiruvananthapuram, Kerala, India 695034
[3]Department of Civil Engineering, College of Engineering, Thiruvananthapuram, Kerala, India 695016
[4]Department of Geology, University College, Thiruvananthapuram, Kerala, India 695034

**Abstract**

We have studied the geological time evolution of volcanism in Earth and other inner solar system planetary bodies (Mercury, Moon, Mars and Venus) in both geophysical and biophysical perspective. The record of LIP's (Large Igneous Provinces) in Earth and other planetary objects suggest the existence of increasing, decreasing and cessation phases of major volcanic activity over geological time scales. We have extended the existing scale of measuring intensity like VEI of volcanic eruptions based on Earth based observations to accommodate intense and extreme volcanic activity. The mass of a rocky planetary object is found to be related to the magnitude of the internal heat, occurrences of LIP's and the duration of major volcanic activity from relevant data available for the inner solar system. The internal heat magnitude may also decide the intensity of volcanism in these planetary objects. The time evolution of volcanism in Earth and Mars has probably influenced the origin of life and biological evolution in these planets.

## Introduction

Volcanism is known to be an important mechanism for the dissipation of internal heat in rocky planetary objects like Earth. Different aspects of volcanism in the inner solar system are studied extensively (Platz et al., 2015; Byrne, 2019). There are models which address the time evolution of volcanism in rocky exoplanets. The phenomenon of cessation of volcanism in planetary objects is not understood in detail so far. Varnana et al (2018) explained the inferred cessation of major volcanism in Moon, Mars, Venus and Mercury in terms of a critical internal heat (comparable to the current surface heat flux of Earth) which is required to sustain the phenomena for rocky planetary objects. They also suggested that the major volcanism in Earth will cease in the near geological future.

In this paper we have studied in detail the geological time evolution of volcanism in the inner solar system. Our investigations yielded the following interesting results and some of them are presented in past conferences.

( i) The magnitude of a particular volcanic eruption series in Earth is measured in terms of total volume ( VEI index) or total mass of eruptions ( M index). The universality of this magnitude scales are still uncertain but are expected to hold well for extra-terrestrial or exo planetary volcanism also. At present VEI scale is defined for eruptions with magnitude ranging from 1-8. In this paper we have extended the VEI scale to include eruptions with magnitude between 9-13.The data for past major ice ages in Earth is also used in this context. A simple conversion procedure from VEI to M scale is also available in the current literature (Mason, 2004)

(ii) A linear relation is found between time variations of maximum intensity of volcanic eruptions in Earth (M) and corresponding internal heat flux (S) variations from best available observations during the past 200 Ma. The geological time evolution of volcanism in Earth is discussed in terms of our simple thermal evolution model and based on the available data on LIP since 3.8 Gyrs ago

(iii) The current internal heat parameters inferred for inner solar system planetary objects is found to be proportional to the basic geophysical properties of these planetary objects such as mass.

(iv)Geological time evolution of volcanism in Moon, Mars, Venus and Mercury will be discussed in terms of our simple thermal evolution model and available observations. The mass-cessation age relations for inner solar system planetary objects will be also investigated in this context.

(v) We have compared geological time evolution of volcanism with the biological evolution in Earth. This suggests the possibility for the existence of advanced life forms along with extreme or high volcanic activity and may be relevant for Mars and potentially habitable extrasolar planets.

## 2. Investigations on the geological time evolution of volcanism on Earth

### 2.1 Extension of the VEI scale of volcanic eruption intensity to study extreme volcanic activity in planetary bodies

The magnitude of volcanic eruptions is at present expressed in terms of volcanic explosivity index (VEI) where the total volume of the material ejected during the particular volcanic eruption sequence ($V_e$) is mainly considered. VEI is a logarithmic index where a change of VEI value by unity (for eg: from 7 to 8) implies a change in $V_e$ by ten times the previous one. Usually VEI is defined up to a magnitude 8 only whose volume in units of km$^3$ .We can find the volume of more intense volcanic eruptions by increasing the value of $V_e$ successively by ten times each as shown in Table 1. The maximum value of $V_e$ is expected to be only a fraction of the total volume of Earth ($10^{12}$ km$^3$). Apart from volume of ejecta the total duration of a particular volcanic eruption sequence ($t_v$) is also important. We can assume that

$$V_e \text{ (km}^3\text{)} = t_v \text{ (years)} \qquad (1)$$

The above assumption is supported by the values of $t_v$ and $V_e$ inferred for several major volcanic eruptions which happened during 250 Myr to 65Myr in the geological past of Earth (Wignall, 2001). In the above cases it is found that $V_e$ is of the order of million km$^3$ and $t_v$ is of the order of million years. But $t_v$ cannot exceed the age of Earth ($t_e$= 4.5 Gyr), so that maximum limit of volcanic intensity for Earth is restricted to VEI=13. The values of $t_v$ and $V_e$ for VEI from 8 to 13 is given in Table 1.

Apart from VEI there is another index defined to measure the intensity of volcanic eruptions called M index. The M index is based on the total mass of the eruptions while VEI index is based on the total volume of the volcanic eruptions. , both are logarithmic scale. A relation existing between Magnitude of Eruption Scale (M) and Volcanic Explosivity Index (VEI) was reported by Mason and others ( Mason et al.,2004).The relationship between two scales of volcanic intensity is also based on the assumption that the density of the volcanic eruption deposits is 1000 kgm$^{-3}$. The conversion from VEI to M scale is given in Table 2 .Here by we have used our described extended scale of VEI described in Table 1.

**Table 1** A simple model for the total volume of volcanic eruption ejecta and total duration of the eruption for Earth and other rocky planets for extending the existing VEI scale to accommodate extreme volcanic events

| Volcanic Explosivity Index (VEI) | Volume of Ejecta (km³) | Duration (years) |
|:---:|:---:|:---:|
| 8 | $10^3$ | $10^3$ |
| 9 | $10^4$ | $10^4$ |
| 10 | $10^5$ | $10^5$ |
| 11 | $10^6$ | $10^6$ |
| 12 | $10^7$ | $10^7$ |

**Table 2 Relationship between Magnitude of Eruption Scale (M) and Volcanic Explosivity Index (VEI) for the deposit bulk volume having densities 1000kgm$^{-3}$**

| Magnitude of Eruption (M) | Volcanic Explosivity Index (VEI) |
|---|---|
| 4-4.9 | 4 |
| 5-5.9 | 5 |
| 6-6.9 | 6 |
| 7-7.9 | 7 |
| 8-8.9 | 8 |
| 9-9.9 | 9 |
| 10-10.9 | 10 |
| 11-11.9 | 11 |
| 12-12.9 | 12 |
| 13-13.9 | 13 |

**2.2 Inferring the intensity of volcanism associated with ice ages in Earth**

In Table 3 we have given the details of four major ice ages in Earth in the geological past. Let us assume that these ice ages are caused by large volcanic eruption sequences then and duration of ice ages is equal of $t_v$ of the volcanic eruption defined by equation (1).Using Table 3.2 we have estimated the VEI corresponding to the volcanism associated with these ice ages whose values are shown in Table 3.

**Table 3 Major ice ages in Earth and inferred intensity of associated volcanic eruptions**

| Ice Age Period in Earth | Duration (million years) | Inferred volcanic intensity in M Scale | Inferred volcanic intensity in the VEI scale |
|---|---|---|---|
| Huronian (2.4-2.1 Gya) | 300 | 13.1-13.9 | 13 |
| Cryogenian (850-630Mya) | 220 | 13.1-13.9 | 13 |
| Saharian (460-420Mya) | 40 | 13.1-13.9 | 13 |
| Karoo (360-260Mya) | 100 | 12.04 | 12 |

**2.3 Evidence for significant decline in volcanic activity in Earth during the past 360 million years**

The past 500 Million years in Earth is important in terms of significant biological evolution which is also associated with prolonged ice ages and mass extinctions. Since the beginning of last major ice age in Earth (Karoo ice age, 360 Myrs ago) we could infer large decline in volcanic intensity in Earth. In Table 4 we have given magnitude (M) of major volcanic eruptions in chronological order.

**Table 4 Details of very intense volcanic eruptions in the past 400 Myrs in Earth**

| Geological Timeperiod (Mya) | Details of Mass Extinction | Details of Ice Age | Inferred Surface Heat Flux (W/m$^2$) | Intensity of Volcanism in M Scale | Reference | Intensity of Volcanism in VEI scale |
|---|---|---|---|---|---|---|
| 360-260 | End Permian | Karoo ice age (360-260 Myr) | 0.1051 (0.1032) | 12.45 | From Table 3 | 12 |
| 64.8 | End Triassic | | 0.0958 | 9.4 | Self et al., 2008 | 10 |
| 27.2 | | | 0.0942 | 9.2 | Lipman ,1997 | 8 |
| 16.5 | | | 0.0938 | 8.8 | Reidel and Tolan,1992; Reidel,1983 | 8 |
| 15.6 | | | 0.0937 | 8.8 | Reidel ,2005; Landon and Long ,1989 | 8 |
| 14.5 | | | 0.0936 | 8.7 | Tolan et al.,1989 | 8 |

**Table 4: Continued**

| Geological Timeperiod (Mya) | Details of Mass Extinction | Details of Ice Age | Inferred Surface Heat Flux (Wm$^{-2}$) | Intensity of Volcanism in M Scale | Reference | Intensity of Volcanism in VEI scale |
|---|---|---|---|---|---|---|
| 6.5 | | | 0.0933 | 8.6 | Morgan et al.,1984 | 8 |
| 0.6 | | | 0.0930 | 8.55 | Christiansen,1984, 2001 | 8 |
| 0.0265 | | | 0.0930 | 8.1 | Wilson, 2001 | 8 |

For the period between 65 million years to recent times we have good coverage of volcanic data and the magnitude M of eruptions is estimated from both mass and volume of eruptions. For this period we could find a linear regression relation between M and surface heat flux of Earth found from our thermal evolution model given by:

$$M = 390.2 \times S - 27.83 \quad [R^2 = 0.78] \qquad (2)$$

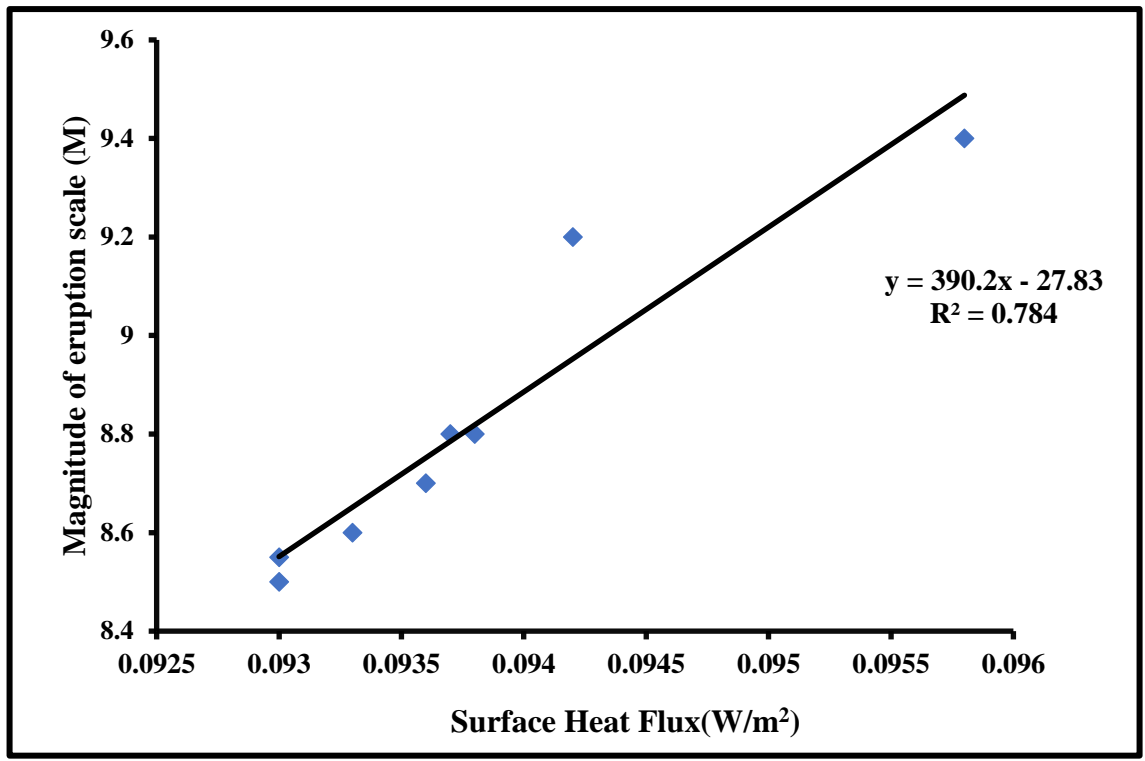

**Fig.1 Linear Regression relation between magnitude of volcanic eruption intensity (in M scale) and inferred surface heat flux of Earth during the past 65 million years**

The calculated value of S for the Karro ice age (360-260 Myrs) from equation (2) by using the value of M of the related eruption is given in brackets in Table 4. The difference between model inference of S and volcanic inference of S of Earth is found to be small. During the past 360 Myrs we can find that the maximum intensity volcanism in Earth has decreased by a factor 5.

**2.4 LIP data and geological time evolution of volcanism in Earth**

A compilation of information about all LIP's in Earth known till date is available (www.largeigneous.com).The earliest LIP is inferred to have occurred 3.79 Ga ago and the recent one is inferred to have occurred 20 million years ago. Details about 215 LIP's are given in this compilation. For most of the LIP's total area or volume of volcanic eruptions are given along with inferred geological age. Using our extended VEI scale of volcanic eruption intensity we have identified those LIP's whose VEI is at least 9.We could find about 111 LIP's with VEI values greater than or equal to 9 from the available LIP record. The results are given in Fig 2 -4. In Fig 2 we have plotted the number of LIP's with VEI between 9-10 found for different periods such as 3 Gyrs, 2.5-3 Gyrs etc. In Fig 3 we have plotted the same with VEI values 10-11 and in Fig 4 the same with VEI values between 11-12 and in Fig 5 the number of LIP's with VEI values between 9-12 is given.

If volcanic activity in Earth had followed the internal heat evolution with time we expect a decrease in number of LIP's with geological time since planetary formation. In contrast to the same we can find from Fig 2-4 that this number is initially low (around 3 Gyrs) and increases to a maximum value about 500-1000 Ma in the geological past and then declines significantly and reaches a value comparable to ancient Earth in the immediate geological past. The inferred internal heat evolution (Varnana et al, 2018) and evolution of maximum volcanic eruption magnitude in Earth is summarized in Fig 6.

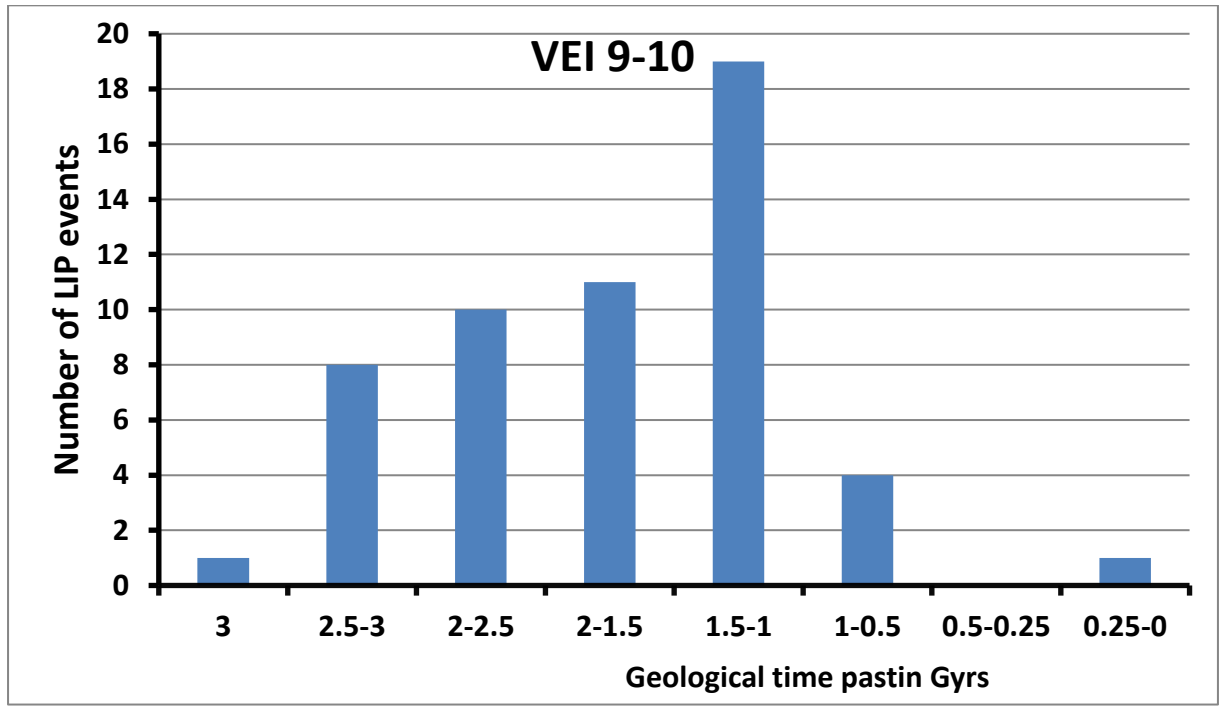

**Fig 2 Number of LIP in different geological ages with inferred VEI between 9-10**

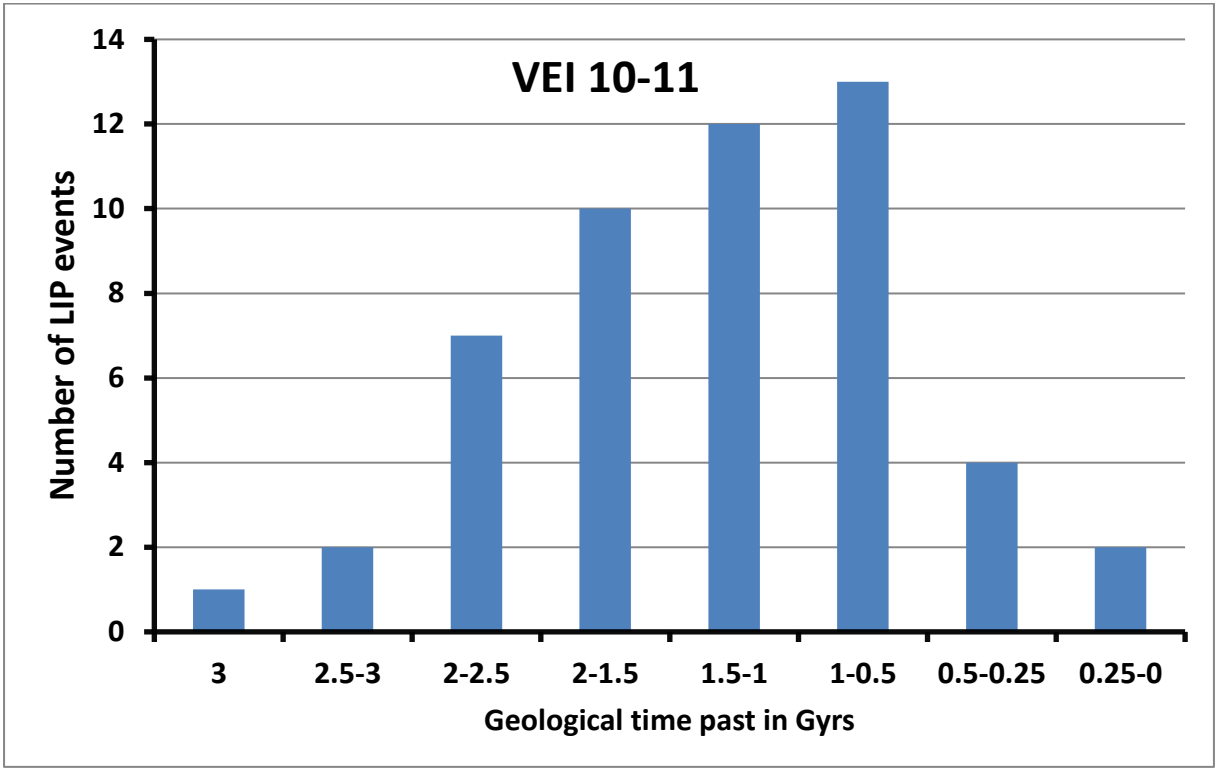

**Fig 3 Number of LIP in different geological ages with inferred VEI between 10-11**

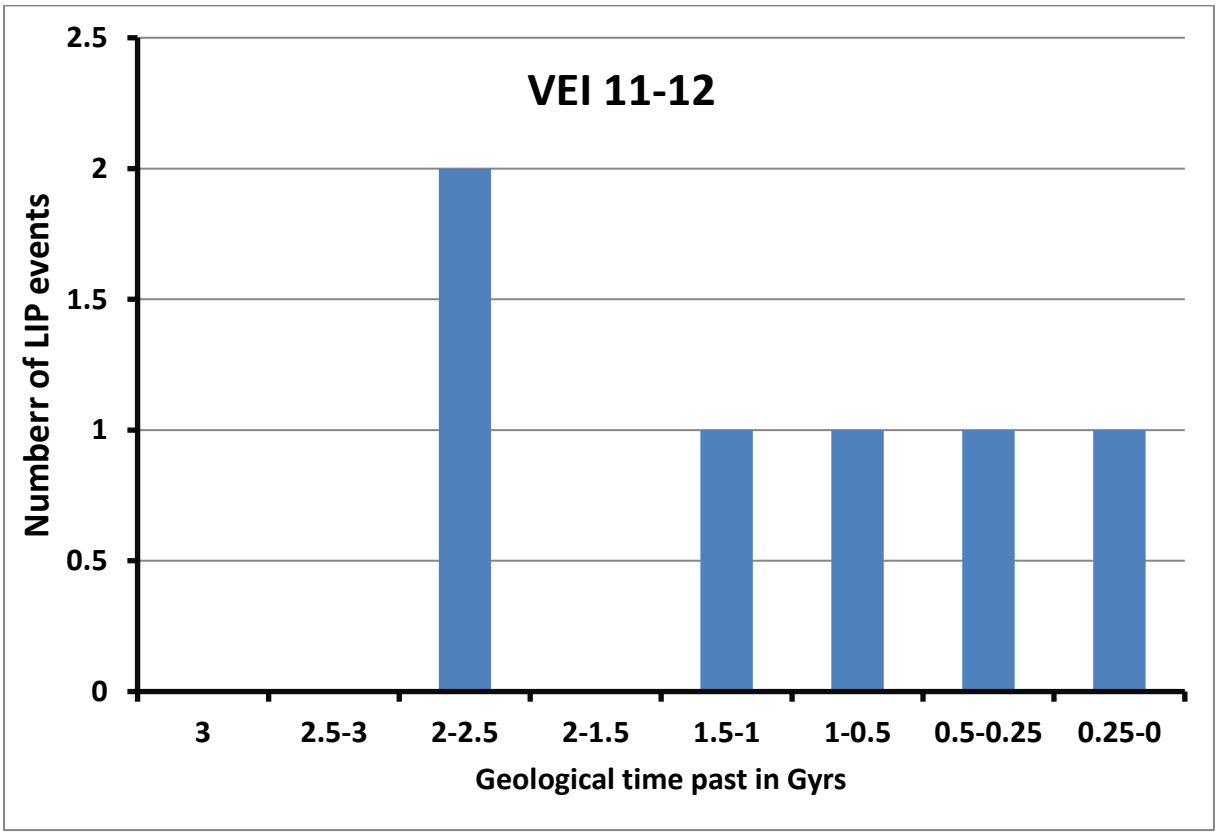

**Fig 4 Number of LIP in different geological ages with inferred VEI between 11-12**

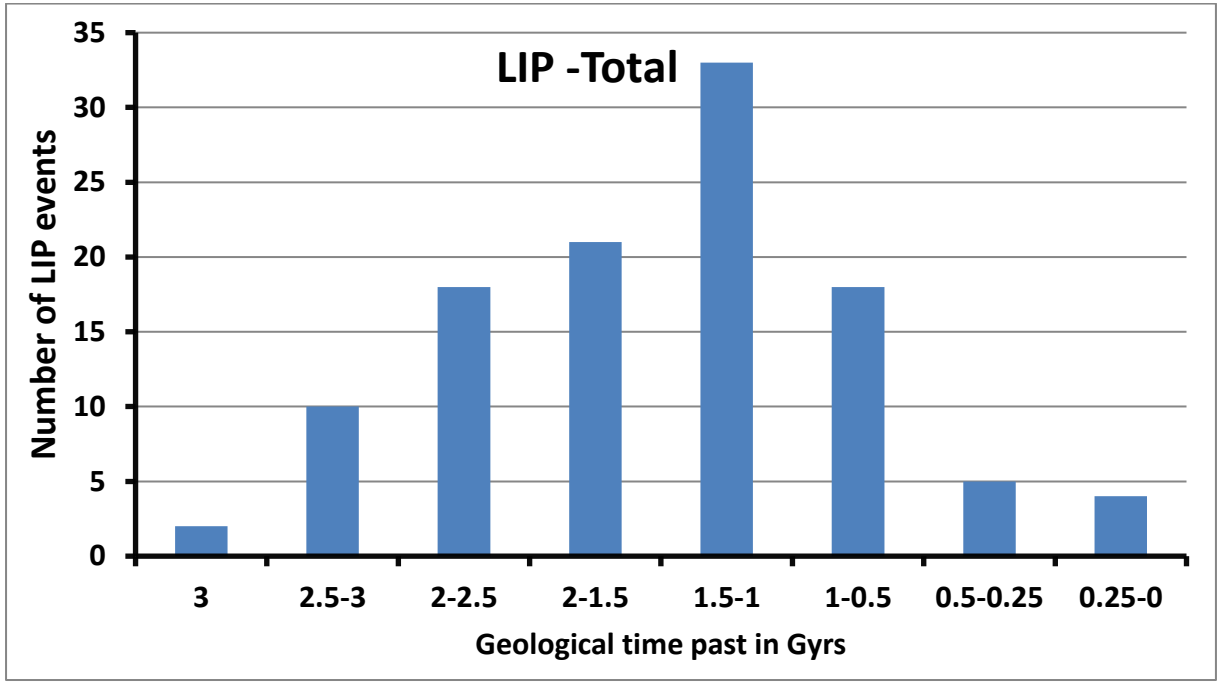

Fig 5 Number of LIP in different geological ages with inferred VEI between 9-12

**3 Geophysical parameters of rocky planetary objects in the inner solar system and volcanism**

**3.1 Finding linear regression relations between surface heat flux and different geophysical parameters of rocky planets in the inner solar system**

Mass (kg), diameter (km), acceleration due to gravity (ms$^{-2}$), density (kgm$^{-3}$) are some of the basic geophysical parameters of the rocky planets in the solar system as given in Table 5.

**Table 5 Basic Geophysical parameters of inner solar system rocky planetary bodies**

| Planet/ Satellite | Mass $10^{24}$ (kg) | Diameter $*10^3$ (m) | Density (kgm$^{-3}$) | g (ms$^{-2}$) | Mantle Depth (km) |
|---|---|---|---|---|---|
| Earth | 5.9726 | 12742 | 5514 | 9.8 | 2890 |
| Venus | 4.8876 | 12103 | 5243 | 8.9 | 2745 |
| Mars | 0.6417 | 6779 | 3933 | 3.7 | 1698 |
| Mercury | 0.3301 | 4879 | 5427 | 3.7 | 618 |
| Moon | 0.0735 | 3474 | 3340 | 1.6 | 1340 |

We can look for linear relationships between surface heat flux S (Wm$^{-2}$) of rocky planets in the solar system and the above geophysical parameters. A general linear regression relation shall be of the form

$$S = Ax + B \qquad (3)$$

Here x is the value of the geophysical parameter of the rocky planet, A is the slope and B is the y intercept.

The estimated values of A, B and $R^2$ values for different geophysical parameters of the rocky planetary objects in the solar system is given in the Table 6. The linear regression best fit plot between planetary masses and their surface heat flux values is shown in Fig 6. The linear regression relation between mass and surface heat flux of planetary body is as follows

$$S = 0.084 \times Mp(EU) + 0.014, \quad R^2 = 0.98 \qquad (4)$$

where

Mp is the mass of the planetary body in Earth mass units

S is the surface heat flux in Wm$^{-2}$ corresponding to the planetary body

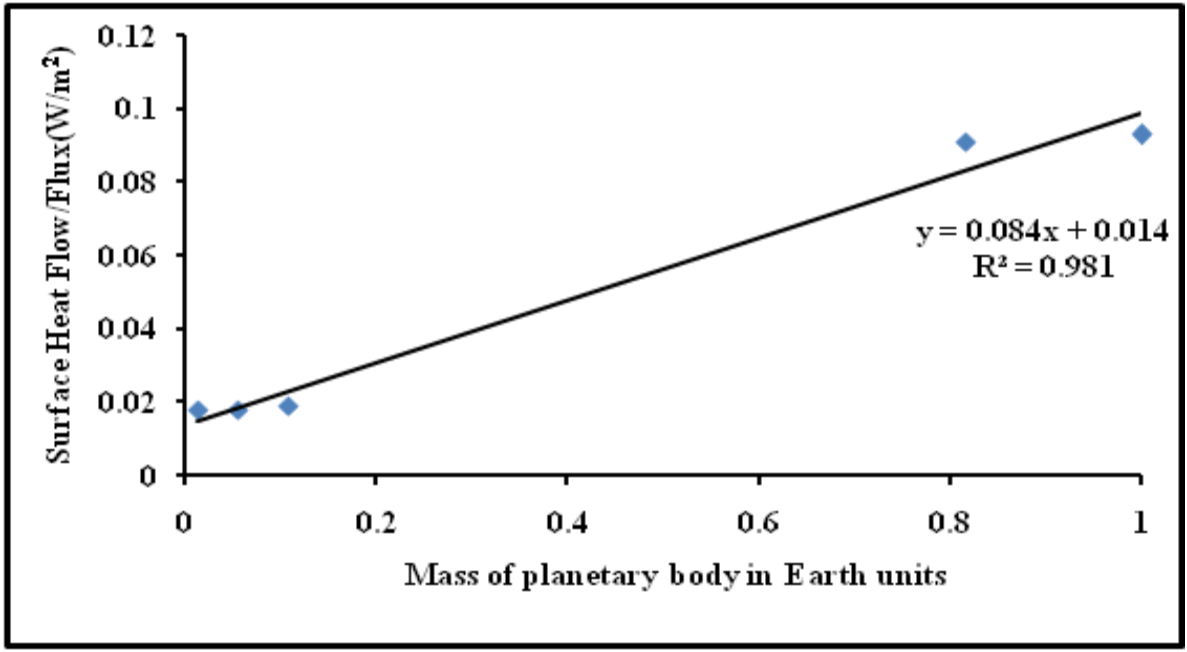

**Fig. 6 Linear regression plot between mass and surface heat flux of rocky planetary objects in the solar system**

**Table 6 The linear regression fit parameters found between the internal heat (Surface heat flux) and other basic geophysical parameters of the inner solar system rocky planetary objects**

| Geophysical Parameter | Slope | Y Intercept | $R^2$ value |
|---|---|---|---|
| Mass | 0.084 | 0.014 | 0.98 |
| Diameter | $9 \times 10^{-6}$ | -0.026 | 0.93 |
| Acceleration due to gravity | 0.010 | -0.012 | 0.94 |
| Density | $3 \times 10^{-5}$ | -0.073 | 0.98 |
| Mantle depth | $4 \times 10^{-5}$ | -0.023 | 0.84 |

We could find a linear relation between mass and internal heat of rocky planetary bodies in the inner solar system which may have significance in planetary astrophysics. Published theoretical models have suggested that internal heat dynamics which drive tectonic or volcanic phenomena in rocky planets may depend on planetary mass (Kite et al., 2009; Noack and Breuer, 2014; Papuc and Davies, 2008). In Fig 7 we have shown a linear regression plot between mass and mantle depth of Earth, Moon, Mars, Mercury and Venus. This relation suggests that mass plays possibly an important role in the mantle dynamics of rocky planetary objects. The tidal heating of solar system rocky planetary bodies by the Sun is calculated and is found to be negligible.

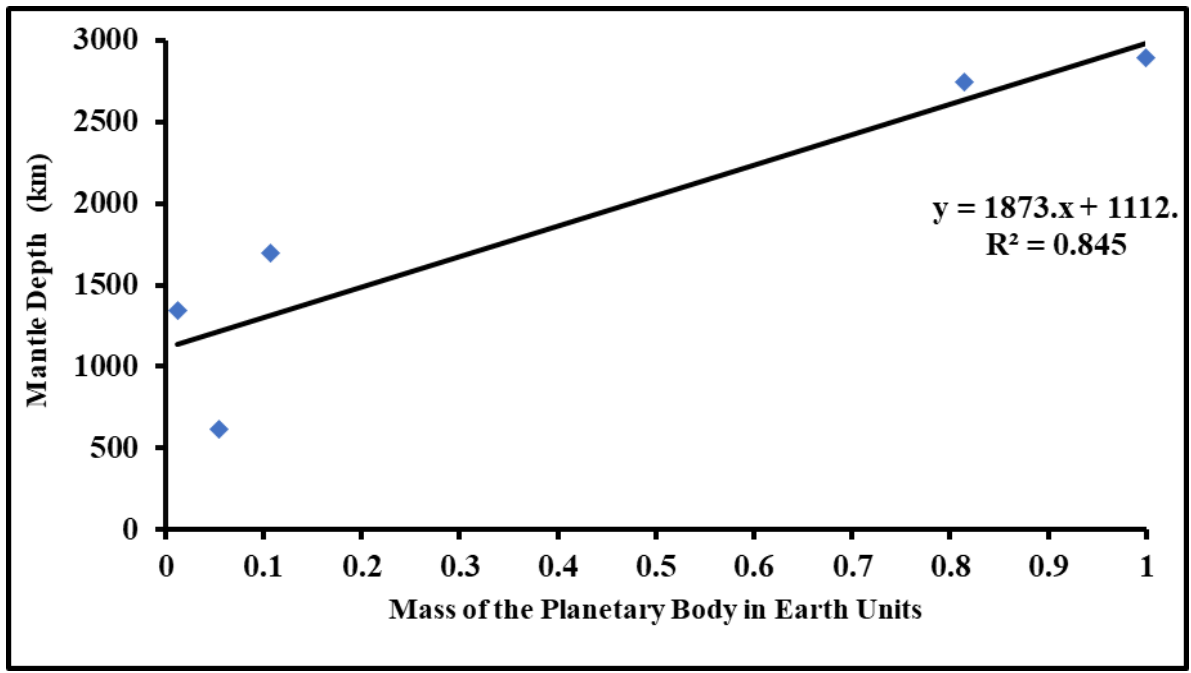

**Fig 7 Relation between mass and mantle depth of rocky planetary bodies in the inner solar system.**

## 3.2 Planetary mass-volcanic cessation age relations

In our model we are suggesting that the surface heat flux of a planetary body varies exponentially with time as given by the relation

$$S = S_0 \exp(-\Lambda t) \qquad (5)$$

But according to (1-4)

$$S = 0.084 \times M_p (EU) + 0.014 \qquad (6)$$

This implies

$$S_0 \exp(-\Lambda t) = 0.084 \times M_p (EU) + 0.014 \qquad (7)$$

At the Earth age the factor $\exp(-\Lambda t)$ becomes constant

Hence

$$S_0 \times constant = 0.084 \times M_p (EU) + 0.014 \qquad (8)$$

We here found out $S_0$ by varying the mass from 1 Earth mass units to 10 Earth mass units.

Once the value of $S_0$ is known, the cessation age ($t_c$) in Gigayear can be calculated by using the relation

$$t_c (Gyr) = 1/\Lambda \ \ln(S_0/S) \qquad (9)$$

The value of decay constant is $1.5 \times 10^{-17}$ s$^{-1}$ (Wood, 1988). The plot of mass versus cessation age is given in the Figure 8

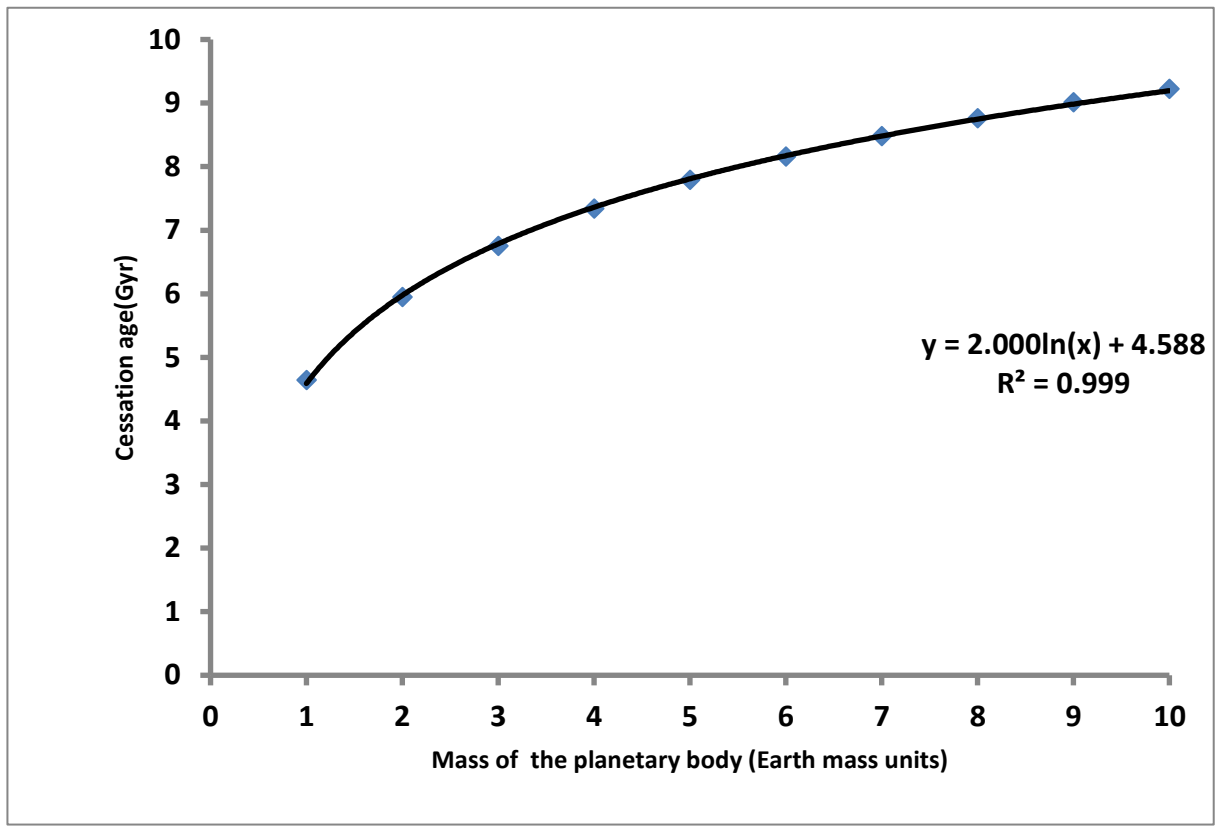

Fig 8: The plot of mass versus cessation age by the variation of mass upto 10 Earth units

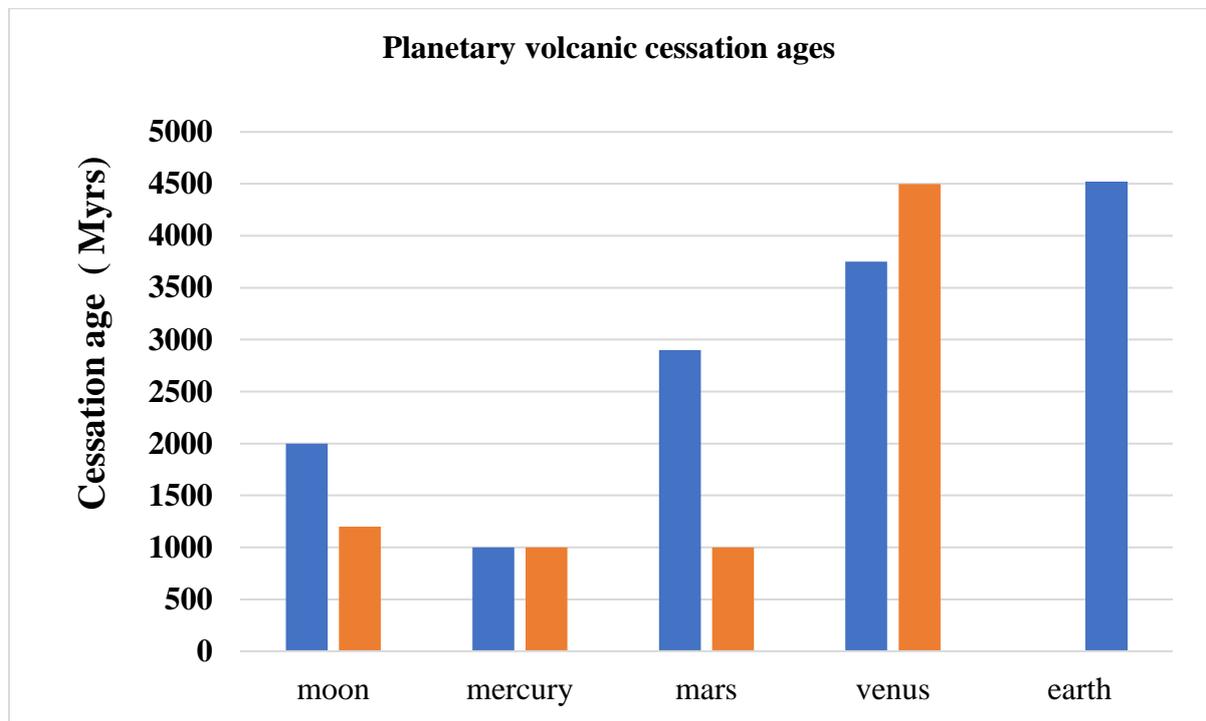

Fig 9 Cessation age of major volcanism (reckoned from formation of planetary system) inferred by Varnana et al (2018) in orange and Byrne (2019) in blue

The cessation age is shorter for less massive planets (Moon, Mercury and Mars) and longer for more massive planets (Venus and Earth) supporting the mass-cessation age relation described in this section.

**4 Geological time evolution of Volcanism in other planetary objects in inner solar system**

Major volcanism once existed in other planetary objects in the inner solar system (Moon, Mars, Mercury and Venus) has already inferred to have ceased (Varnana et al, 2018) .It will be interesting to know how major volcanism evolved in these planetary objects during the geological past. We do not have long LIP record like Earth for these objects (Ernst, 2014) and hence our knowledge will be definitely incomplete. In Table 7 we have given details of period of intense volcanism and maximum inferred VEI for these planetary objects. The total number of LIP's observed during the geological past for each planetary body including Earth is also given in this Table 7. The values of volcanic eruption intensity during the cessation period and after are also given if available. The internal heat evolution along with known/inferred volcanism related milestones for different planetary objects in the inner solar system is shown in Fig 10-13.

**Table 7 Details of inferred/observed of volcanic intensity in different planetary objects during selected geological ages**

| Planetary body | Period of peak volcanism in the past and inferred VEI | Inferred S value during peak volcanism (Wm$^{-2}$) | Number of LIP's during geological past | Reference | Observed VEI of present or recent volcanic activity |
|---|---|---|---|---|---|
| Moon | 3.3-3.9 Gyrs VEI: 10-11 | 0.0856-0.1137 | 8 | Spudis et al., 2013 | VEI < 7 |
| Mercury | 3.55-4.1 Gyrs VEI : 10-11 | 0.0963-0.1249 | 9 | Byrne et al.,2016 | NA |

**Table 7 Continued**

| Mars | 3.7-4.1 Gyrs VEI: 12-13 | 0.1089-0.1316 | 20-28 | Robbins et al.,2011 | VEI < 7 |
|---|---|---|---|---|---|
| Venus | 0.3-1 Gyrs VEI : 9-12 | 0.1048-0.1459 | 168 | Romeo and Turcotte 2009 | VEI:7 |
| Earth | 1-0.5 Gyrs VEI; 10-12 | 0.1177-0.1492 | 111 -200 | www.igneousprovinces.org | VEI< 7 |

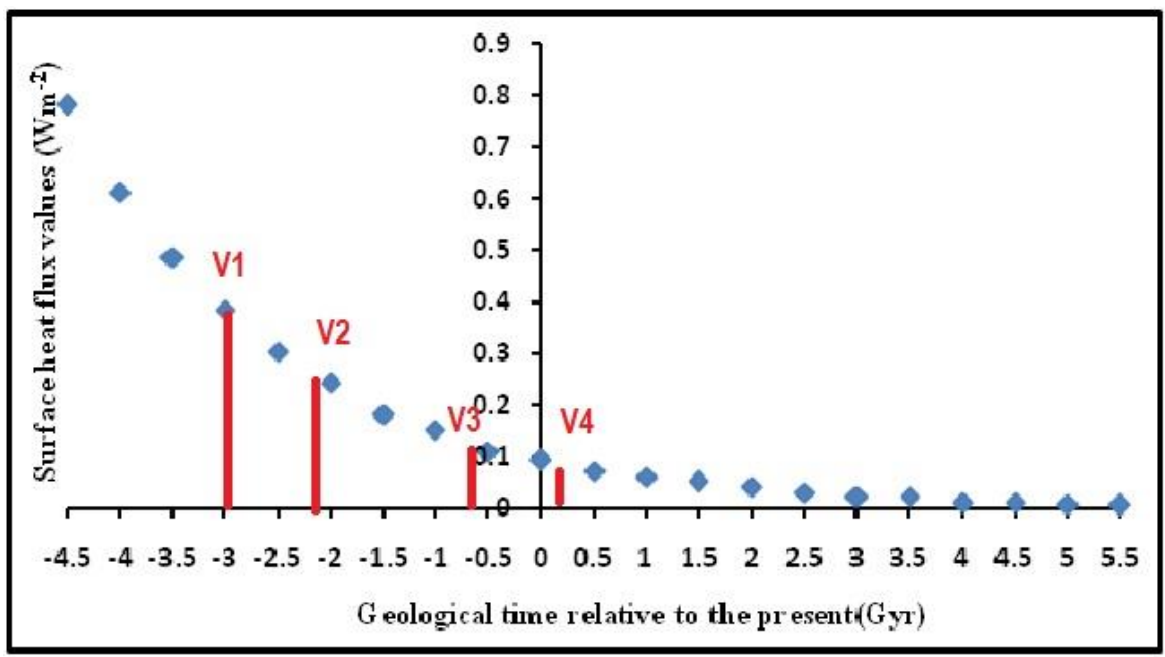

**Fig 10: Geological time evolution of internal heat in Earth**

(**V1-Starting of intense volcanism, V2-Extreme volcanism, V3-Peak of LIP volcanism, V4-Cessation of volcanism**)

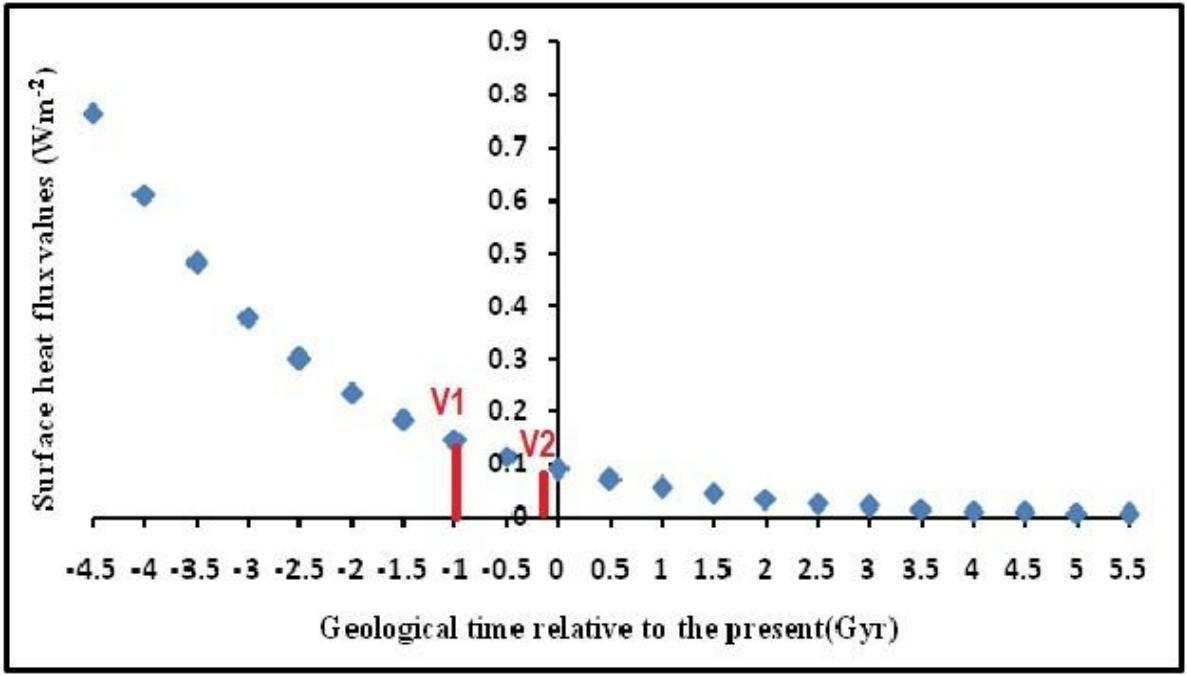

**Fig 11: Geological time evolution of internal heat in Venus**

**(V1-Beginning of peak volcanism, V2-Volcanic cessation age)**

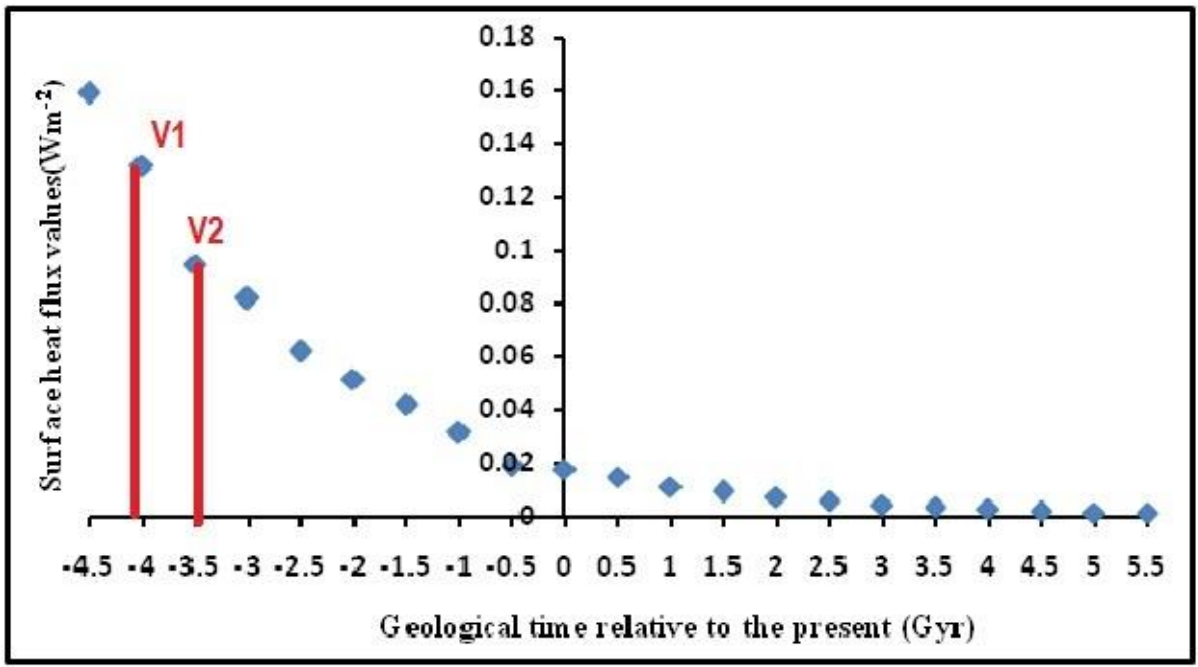

**Fig 12: Geological time evolution of internal heat in Mars**

**(V1-Beginning of peak volcanism, V2-Volcanic cessation age)**

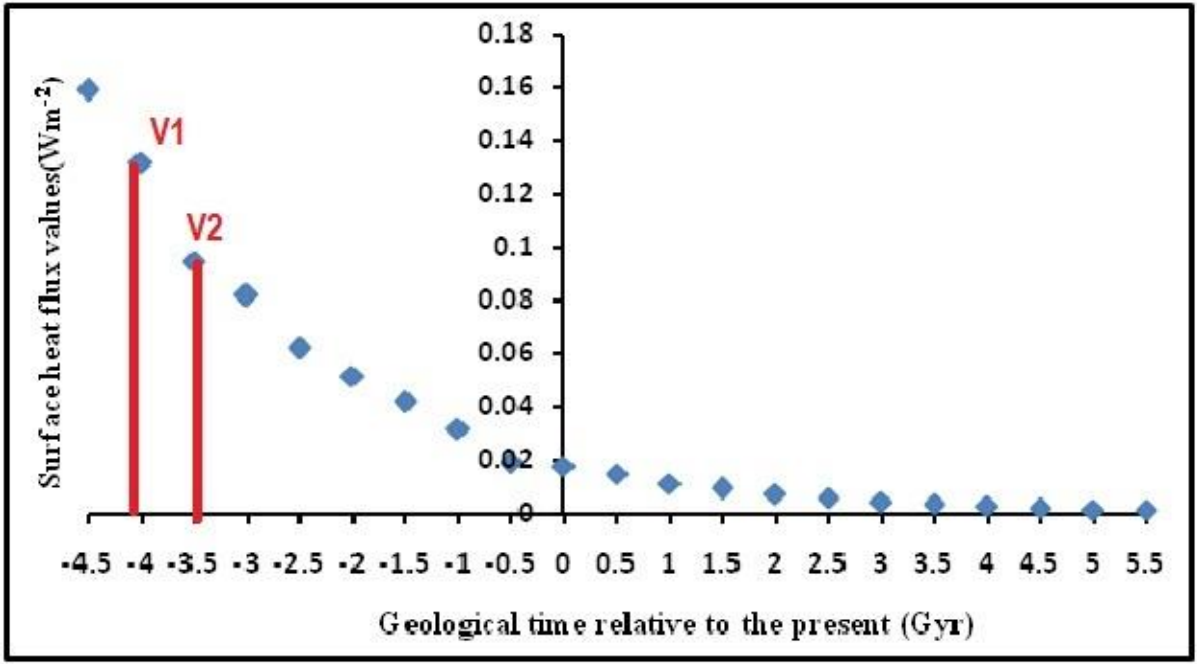

**Fig 13: Geological time evolution of internal heat in Mercury**

**(V1-Beginning of peak volcanism, V2-Volcanic cessation age)**

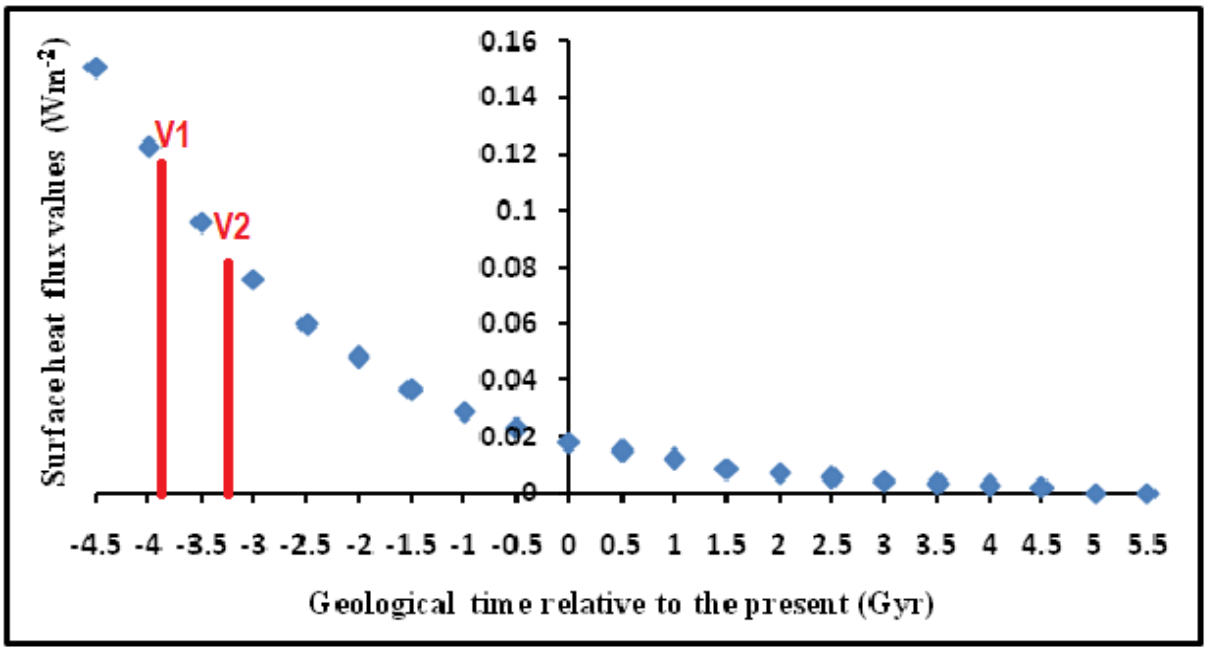

**Fig 14: Geological time evolution of internal heat in Moon**

**(V1-Beginning of peak volcanism, V2-Volcanic cessation age)**

## 3.5 Volcanism and biological evolution in Earth

In Table 8 we have given important mile stones of the biological evolution in Earth along with inferred surface heat flux and intensity of volcanism associated with each period. Six major mass extinctions happened during the geological past of Earth (Erwin, 1998) of which the first five are associated with major ice ages (see Table 3). Ice ages are related to intense volcanic eruptions or LIP's. The first mass extinction is associated with snow ball earth or Cryogenic ice age period (650 Mya) (Hoffmann and Schrag, 2000) and the last one is associated with the Deccan trap volcanism in NW India which led to KT mass extinctions (Glasby and Kunzendorf, 1996).In Fig 10 we have shown the inferred internal heat values (surface heat flux, S) of Earth during these mass extinction periods. We can find a systematic decrease in S from the first to the last mass extinction periods which imply a systematic change in volcanic activity in Earth .We can also infer that between 260 Myrs (end of Karro ice ages) to Deccan trap period (65Myrs) (Curtilot,1988) , the surface heat flux of Earth (S) decreased by 15 % . After the KT extinction we can observe significant development of advanced and intelligent life in earth. The evolution of mammals, primates and apes happened during this epoch which finally led to the emergence of modern humans. During this post K-T extinction period we can find that the intensity of volcanism is moderate (VEI: 7-8).

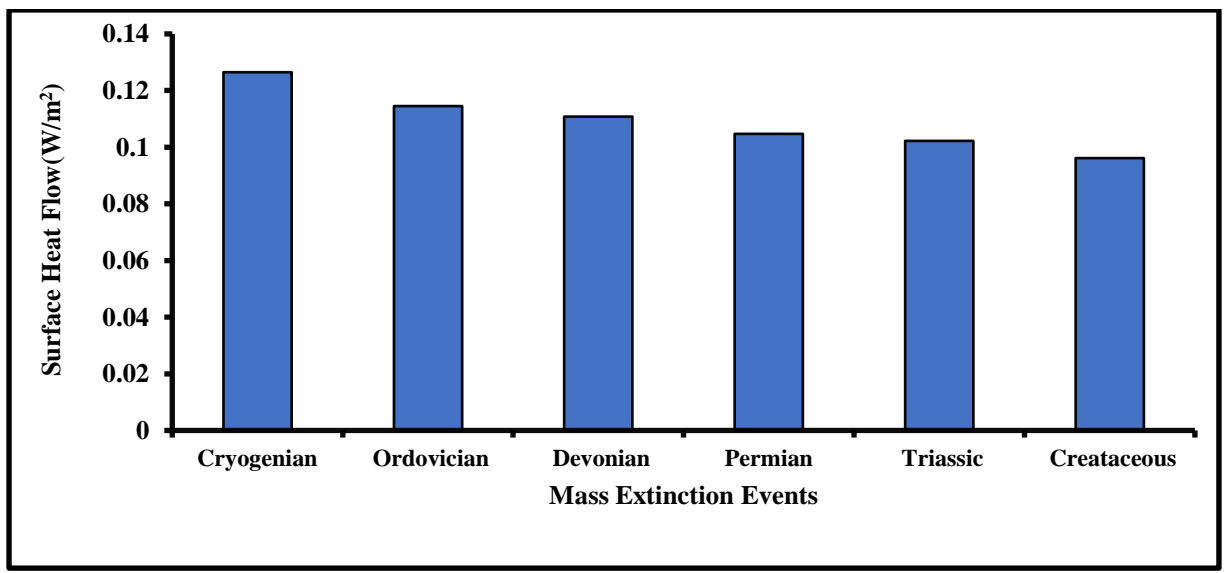

**Fig15 Inferred Surface Heat Flow values during the major mass extinction events in Earth**

**Table 8 Important milestones of biological evolution in Earth and inferred surface heat flux and volcanic eruption intensity during these periods**

| Timeline | Biological event | Surface heat flow/flux (Wm$^{-2}$) | Max intensity of volcanism (VEI) |
|---|---|---|---|
| 3.8 billion years ago | bacteria and archaea. | 0.5613 | ---- |
| 3.5 billion years ago | Single-cell organisms. | 0.4870 | --- |
| 3 billion years ago | Viruses | 0.3844 | 10-11 |
| 2.4 billion years ago | great oxidation event | 0.3035 | 10-11 |
| 2.15 billion years ago | Cyanobacteria and photosynthesis | 0.2572 | 10-11 |
| 900 million years ago | first multicellular life | 0.1424 | 10-11 |
| 590 million years ago | Animals with bilateral symmetry, divide into the protostomes and deuterostomes | 0.1229 | 10-11 |
| 580 million years ago | jellyfish, sea anemones and corals | 0.1223 | 10-11 |
| 540 million years ago | first chordates | 0.1200 | 10-11 |
| 520 million years ago | earliest vertebrates | 0.1183 | 10-11 |
| 465 million years ago | Land plants | 0.1156 | 10-11 |
| 400 million years ago | insects | 0.1124 | 10-11 |

**Table 8 Continued**

| | | | |
|---|---|---|---|
| 397 million years ago | first four-legged animals or tetrapods | 0.1122 | 10-11 |
| 385 million years ago | trees | 0.1116 | 10-11 |
| 340 million years ago | amphibians | 0.1092 | 11-12 |
| 310 million years ago | Early mammals. | 0.1076 | 11-11 |
| 210 million years ago | Early birds | 0.1027 | 10-11 |
| 200 million years ago | Dinosors | 0.1022 | 11-12 |
| 150 million years ago | Archaeopteryx famous "first bird" | 0.0998 | 10-11 |
| 130 million years ago | first flowering plants | 0.0989 | 11-12 |
| 75 million years ago | Primates | 0.0964 | 10-11 |
| 70 million years ago | Grasses | 0.0963 | 9-10 |
| 65 million years ago | The Cretaceous-Tertiary (K/T) extinction .The extinction clears the way for the development of mammals | 0.0959 | 9 |
| 63 million years ago | Primates split into two groups, known as the haplorrhines and the strepsirrhines | 0.0958 | 9 |
| 40 million years ago | New World monkeys | 0.0947 | 9 |

**Table 8 Continued**

| 25 million years ago | Apes split from the Old World monkeys | 0.0941 | 8 |
|---|---|---|---|
| 18 million years ago | Gibbons become the first ape to split from the others. | 0.0938 | 8 |
| 14 million years ago | Orang-utans branch off from the other great apes, | 0.0936 | 8 |
| 7 million years ago | Gorillas branch off from the other great apes. | 0.0933 | 8 |
| 6 million years ago | Humans diverge from their closest relatives; the chimpanzees and bonobos | 0.0932 | 8 |

**Discussion and Results**

We have used the long LIP record available for Earth to study the geological time evolution of volcanic activity in our planet. This is not done explicitly in previous studies. Varnana et al (2018) distinguished major volcanism from minor volcanism for rocky planetary objects in the inner solar system and inferred that the major volcanism in Earth will cease in the near geological future. Byne (2019) also inferred the cessation periods of major volcanism for different planetary objects in the inner solar system from the available records. But in his paper the intensity or magnitude evolution of volcanism is not shown explicitly. From the present study we could find several interesting results on the geological time evolution of volcanism in the inner solar system planetary objects. This will be discussed below.

We have extended the present magnitude scale of volcanic eruptions to include both intense and extreme volcanic activity found for earth and similar rocky planetary objects. In the VEI scale there is an apparent relation between total duration and total volume of a given series of volcanic eruptions. This is applied to available LIP record of earth to identify intense and extreme volcanic eruptions in Earth in the geological past. The conversion from VEI (total volume of eruptions) to M (total mass of eruptions) scale of magnitude of volcanic eruptions is also discussed. Our extended magnitude scale is also used to study intense volcanism in other planetary objects in our inner solar system

The LIP records of Earth (this can be considered as an important manifestation of major volcanism) suggest that intense volcanic activity in our planet began only around 3 Gyrs ago. The number of very intense volcanic eruptions (VEI: 10-11) is found to increase gradually with geological time and reaches a peak between 1- 0.5 Gyrs ago and then declines significantly. This decline supports the inference of Varnana et al (2018) about the possibility of cessation of major volcanism in Earth in the near geological future.

The extreme volcanic activity as represent by LIP eruption events with highest magnitude (VEI 11-12) shows a remarkable association with the periods of major ice ages in Earth ( Table 3 ) . The first extreme magnitude LIP events happened between 2.1-2.2 Gyrs ago. This coincides with the Huronian ice age in Earth between 2.1-2.4 Gyrs ago. During the second ice age period (snowball Earth) between 850-630 Ma ago also there is an extreme volcanism event during 727 Ma ago. Close to the beginning of last major ice age (360 Ma-Karro ice age) we have an extreme magnitude LIP dated 365 Ma ago.

We could find linear regression relations between basic geophysical parameters such as mass and current internal heat parameters of inner solar system planetary objects. Planetary mass possibly decides its thermal and volcanic history over geological time scales. In this context we could find a planetary mass-volcanic cessation age relation theoretically. This relation is supported by the inferences of cessation ages of inner solar system planetary objects as discussed in section. For the case of Earth we have also found a linear regression relation between surface heat flux and maximum intensity of volcanism observed/inferred using relevant observations for the past 200 Ma. The number of LIP's observed during the geological history of a planetary body seems to depend on its mass if we examine Table 1. In the inner solar system for the most massive planets: Earth and Venus, the above number is significantly higher than its companions. The number observed during the geological history of Mars, Mercury and Moon decrease in the order of their respective mass. Mass-internal heat relations and internal heat-volcanism relations discussed in this paper can justify this result

In section we have discussed geological time evolution of the volcanic eruption intensity for Moon. Mars, Mercury and Venus. In the case of inner solar system planetary objects we could identify three phases of volcanic evolution over geological time scales .They are (i) increasing phase of volcanic activity and reaching a peak (ii) this followed by a decreasing phase of volcanic activity and cessation of major volcanism (iii) Post cessation minor volcanism (ii) Moon and Mars have completed phase (i) and (ii) in the remote geological past and is now in phase (iii). Mercury is also similar to these planets but not entered phase (iii). Venus has completed phase (ii) in the recent geological past and yet to enter phase (iii). Earth is inferred to be close to the end of phase (ii) and we are yet to know whether this will be followed by phase (iii). One interesting result is that from detailed analysis of LIP record of Earth we could infer a clear asymmetry between duration of phase (i) and phase (ii). Duration of phase (i) for Earth is found to be significantly longer than phase (ii). The former is having a value of more than 2.5 Gyrs and the latter is inferred to have a value of only 220 Ma. However this asymmetry is not expected for low mass planetary objects in the inner solar system like Moon, Mars and Mercury. For Venus we can expect an asymmetry even though our knowledge of Venus volcanism in the remote geological past is still uncertain.

It appears that advanced life in Earth (land plants and mammals) emerged since 500 Ma ago after the peak period of volcanic activity in Earth as known from LIP record. Biological evolution is associated with major mass extinctions which in turn are related to intense volcanic eruptions and ice ages. Intelligent life emerged only after significant decline of volcanic eruption intensity in Earth in the recent geological past.

Primitive life in the form of micro organisms like bacteria possibly existed in Mars during the epoch of peak volcanic activity (3.7-4 Gyrs ago) as shown in Table 8. During this epoch we could also infer co existence of large water bodies and magnetic fields which can shield harmful radiation from space (Lammer et al., 2002) .Oxygen is also recently detected in Mars and our knowledge of oxygen in ancient Mars atmosphere is uncertain (Sholes et al., 2017). After the cessation of major volcanism in Mars around 3.5 Gyrs ago (Varnana et al, 2018) habitability conditions in Mars possibly reversed which includes disappearance of planetary scale magnetic fields and drying up of water bodies. If geophysical changes in Mars repeat in Earth after possible cessation of major volcanism in the near geological future there will be also serious threat to life in our planet.